# Resistivity Ratio of Niobium Superconducting Cavities*

E. L. Garwin and Mario Rabinowitz

*Stanford Linear Accelerator Center, Stanford University, Stanford, California 94305*

Inquiries to: *Armor Research*
*715 Lakemead Way, Redwood City, CA 94062*
*Mario715@earthlink.net*

**Abstract**

Resistivity measurements have been made on Nb cavities, as well as on Pb and Cu, at 296, 77, and 4.2 °K by means of a contactless induced-current method. For superconductors, a constant magnetic field drives the material normal below the transition temperature. These measurements provide a simple means for initial material evaluation as well as a direct means of monitoring the effects of material parameters (purity, heat treatment, gas incorporation, etc.) on the electron mean free path. Approximate determinations of $H_c$, $H_{c1}$, and $H_{c2}$ can also be derived from these measurements. Normal-state thermal conductivity and the Ginzburg-Landau parameter $\kappa$ are calculated from the resistivity measurements.

Measurement of the resistivity ratio (between 296 and 4.2 °K) of Nb rf superconducting cavities is desirable for many reasons. Resistivity-ratio measurements before and after vacuum firing can indicate the decrease in impurity content and other crystalline defects, as well as measure the increase in the electron mean free path. The latter parameter is important in designing superconducting accelerators and in analyzing the results from superconducting cavities. The effects on the electron mean free path due to irradiation, incorporation of gases, etc., can also be determined. The difficulty in measuring resistivity by ordinary means lies in the size and unwieldy geometry of a cavity. To date, no resistivity-ratio measurements have been reported on Nb cavities or even on representative samples, though many speculations have been made regarding electron mean free path in cavities.

We have used a contactless induced-current method for measuring resistivity similar to that used by others [1,2] for normal metals, with the addition of a constant magnetic field to drive the superconductor normal. As far as we know, our measurements are the first using this method on superconductors below their transition temperature, and on actual superconducting cavities.

With a constant magnetic field parallel to the pulsing field, the induced currents in the sample are perpendicular to the constant field, and the radially directed Lorentz force has no net force on samples having origin-reflection symmetry cross sections. However, there is an electrical coupling through the mutual inductance between the constant field coil and the pulsing field coil, which affects the decay time of the induced currents. We found that for practical purposes this cannot be made negligible.

A constant magnetic field applied perpendicular to the pulsing field eliminates the mutual-inductance coupling. However, now the Lorentz force produces a torque on the sample, tending to rotate the axis of the sample into parallelism with the constant field. Mechanical motion caused by this torque produces a time rate of change of flux through the pickup coil, causing a noise signal to be superimposed on the decaying induced-current signal. Because of mechanical support constraints, the sample will oscillate about its equilibrium position, producing an oscillatory superimposed noise signal. The mechanical constraints can be made sufficiently rigid to minimize the oscillation amplitude, and hence the noise.

A small secondary pickup coil of N turns is place around the sample conductor in the region where the resistivity $\rho$ is to be determined. A pulsing solenoid encompasses both the pickup coil and the sample. If the solenoid is suddenly de-energized, the time rate of change of magnetic flux induces currents to flow in the sample to oppose this change in flux. From Maxwell's equations, the voltage induced in the pickup coil is

$$V(t) = N \int \frac{\partial B}{\partial t} dA = N \int D\nabla^2 B dA, \qquad (1)$$

where $B$ is the magnetic flux density in the sample and D is the magnetic diffusivity. Solutions to Eq. (1) exist for various geometries. [1] The result they have in common and which should also be true for any geometry is that at large times

$$V(t) \propto N \exp(-t/\tau), \qquad (2)$$

where the decay time $\tau \propto 1/\rho$. Where analytical solutions exist, $\rho$ may be determined directly from $\tau$. In the case of complicated shapes, such as rf cavities, the resistivity ratio between room temperature and temperature T is simply

$$\frac{\rho(296\ °K)}{\rho(T)} = \frac{\tau(T)}{\tau(296\ °K)}. \qquad (3)$$

The resistivity ratio $r = \rho(296\ °K)/P$, is one of the most sensitive determinations of impurities and other lattice defects, particularly for highly pure samples. (The residual resistivity $\rho_o$, occurs for $T \geq 4.2\ °K$, unless the lattice is extremely free of defects.) However, for our purposes, it is not necessary to obtain the absolute value of the resistivity precisely. When this is desired, any one of the following methods may be used to determine $\rho(T)$ from Eq. (3) for complicated geometries. Since the resistivity at 296 °K does not vary greatly from sample to sample of a given metal, one may (i) use either a literature value of $\rho(296\ °K)$, or (ii) make a dc measurement on a representative rod; (iii) when neither is feasible, $\rho(296\ °K)$ may be obtained by comparison to measurements on a sample of a different material but identical geometry and known resistivity, such as OFHC copper. This experimentally gives the geometrical factor relating $\rho$ and $\tau$. Using these three methods, a typical value of Nb resistivity at 296 °K is $1.6 \times 10^{-7}\ \Omega m$. Even for simple geometries, such as cylinders, if the length-to-diameter ratio is less than 4 or so, the one-dimensional analytic solutions do not give the correct relation between $\rho$ and $\tau$.

The shortest τ measured with no sample in the coil was 2.6 x $10^{-5}$ see, whereas the longest was 7.5 see with Pb at 4.2 °K. Fast and reliable switching of the pulsing coil was made possible by construction of a liquid Ga vacuum switch. τ was obtained from either a semilog plot of the V(t) data from the oscilloscope trace or by matching the V(t) curve on a storage oscilloscope with a variable RC pulser which we made for this purpose. There was ± 8% agreement between τ =RC and τ from the semilog plot, as well as between measurements.

Stock material in the form of solid Nb cylinders from which cavities are machined was measured. The resistivity ratio varied from 26 to 83, depending on impurity content and lattice structure, whereas the ratio to 77 °K was approximately 5 for all these samples. A solid cylinder, 4.75 cm in diameter, and 7.6 cm long, having r = 26 was vacuum fired at 1800 T for 16 h and for another 31 h at about 2000 °C at $10^{-9}$ Torr. This increased r to only 36. The initial impurity content for this sample is ~ 1000 ppm with - 300 ppm of Ta and Zr as the major impurities. This is typical of the as-received Nb stock. Since the resistivity measurement senses the entire bulk of the sample, and the decrease in impurity content is governed both by diffusion and surface evaporation, a much larger increase in resistivity ratio would be expected for a hollow cylinder or cavity.

A $TEO_{11}$ X-band cavity was made from this very same stock. It was vacuum fired five times at pressures ~ $10^{-9}$ to $10^{-8}$ Torr under conditions similar to those of the solid Nb cylinder. At this stage, the cavity was found to have r = 143, residual Q of $Q_o$ ~ $10^9$, with a breakdown field, $H_p'$ ~ 150 Oe. It was subsequently refired at 2200 °C for 50 h with a final pressure of 1 X $10^{-9}$ Torr at 2200 °C . After this treatment, r increased to 500, with $Q_o$ = 1. 5 X $10^9$ and $H_p'$= 260 Oe. However, after high-power operation, $Q_o$ deteriorated to 5 X $10^8$. The resistivity ratio may be higher near the surface within the penetration depth, due to a decrease in the concentration of impurities near the surface. However, for this cavity, having a 4.37-mm wall thickness, r near the surface may not differ substantially from the measured bulk value, considering the amount of high-temperature vacuum firing it received.

As has been shown analytically, the Q and the breakdown field of a cavity may be dominated by trapped flux and/or other local irregularities. This is why there may be no simple correlation between r and $Q_o$ and $H_p'$. Nevertheless, the resistivity measurements may be used as a simple means for initial material evaluation wherein substandard material with too high an impurity content can be rejected. As has been pointed out [6], there may even be an ideal normal resistivity, below or above which power losses would be increased in the cavity. Additionally, the normal resistivity is an important parameter in determining magnetic breakdown. [4]

Not only have resistivity measurements not been reported for a cavity or a representative sample, but neither have thermal- conductivity measurements. The normal thermal conductivity

$$K_n = k_{1n} T \tag{4}$$

is derivable from the Wiedemann-Franz law,

$$K_n = (2.45 \times 10^{-8} \text{ W } \Omega/°K^2) T/\rho, \tag{5}$$

at least for the less pure samples. For the previously described solid Nb cylinder, after its final firing, Eq. (5) gives $K_n = (5.5 \text{ W/m }°K^2)/T$, which is reasonable.

Goodman [7] has derived an expression relating the Ginzburg-Landau parameter $\kappa$ to $\rho$:

$$\kappa = \kappa_0 + 2.37 \times 10^6 \gamma^{1/2} \rho, \tag{6}$$

where $\kappa_0 = 1.22$ is the value of $\kappa$ for pure Nb [8] at 4.2 °K, $\gamma \sim 7.4 \times 10^2$ J/ m$^3$ °K$^2$ is the electronic specific heat for Nb [8], and $\rho$ is the resistivity in ohm meters. Using Eq. (6), we find that $\kappa$ varies from 1.62 to 1.24 as tabulated in Table I.

Rosenblum *et al.* [9] relate $H_{c2}$ to $\rho$ on the basis of an empirically modified theoretical expression:

$$H_{c2} = 0.0321\, T_c^2 (1 - t^2)(1.59 - 0.72 t + 0.097\, T_c \rho), \qquad (7)$$

where they have $\rho$ in micro-ohm centimeters, $H_{c2}$ in kilo-oersteds, and $t = T/T_c$. Taking $T = 4.2$ °K and $T_c = 9.1$ °K, we obtain the values given in Table I.

TABLE I. Measured and calculated properties of Nb samples at 4.2 °K.

| Sample | $\gamma$ | $\kappa$ | $H_{c2}$ (R-A-G) (Oe) | $H_{c2}$ (meas.) (Oe) | $H_{c2} = \sqrt{2}\, \kappa H_c$ (Oe) | $H_{c1}$ (Oe) | $l_e$ (Å) |
|---|---|---|---|---|---|---|---|
| Nb1A | 26 | 1.62 | 3770 | 3800 | 3660 | 990 | 600 |
| Nb1B | 36 | 1.51 | 3460 | 3600 | 3410 | 1040 | 830 |
| Nb2 | 46 | 1.44 | 3280 | 3200 | 3250 | 1080 | 1060 |
| Nb3 | 83 | 1.35 | 3000 | ... | 3050 | 1100 | 1900 |
| Nb cavity A | 143 | 1.29 | 2850 | ... | 2920 | 1150 | 3300 |
| Nb cavity B | 500 | 1.24 | 2700 | 2700 | 2800 | 1170 | 11600 |

These calculated values for $H_{c2}$ agree well with the values obtained directly from the resistivity measurements on the solid Nb cylinder, as shown in Table 1. ($H_{c2}$ measurements were not made on samples Nb3 and Nb cavity A.) In the mixed state for Nb, as in the intermediate state for Pb, the decay time $\tau$ varies as a function of the perpendicular magnetic field, $H$. When $H_{c2}$ is exceeded for Nb, $\tau$ is approximately independent of $H$, neglecting a slight magnetoresistive effect. Thus the threshold for constant $\tau$ determines $H_{c2}$.

An experimental determination of the demagnetizing factor allowed a determination of $H_{c1}$. When the sample is totally in the superconducting state, the decay curve is very steep and constant up to $H_{c1}$, corresponding to the small amount of flux between the pickup coil and a penetration depth in the specimen. As soon as flux begins to penetrate the specimen, the decay curve alters and does not break quite as sharply, for both type-I and -II superconductors. Pb served as a calibration to determine the demagnetization factor by using its known $H_c$ value to calculate the field enhancement, and thus determine $H_{c1}$ = 1.1 kOe for the solid Nb cylinder, Nb1B. This agrees well with the value of 1040 Oe obtained from $\kappa$ determined in our experiments, and the graph of $\kappa$ vs $H_{c1}$ in Harden and Arp's paper. [10] $H_{c2}$ has been determined previously by others [9] using ordinary dc resistivity measurements. This is the first time that either $H_{c1}$ or $H_{c2}$ has been determined by induced-current resistivity measurements.

The electron mean free path $l_e$ can be obtained from the resistivity measurements. Niobium has a complicated Fermi surface, and $l_e$ varies with position on this surface. Goodman and Kuhn [11] have derived the following expression for the average value of $l_e$ taken over the Fermi surface of Nb:

$$l_e = 3.7 \times 10^{-12} \, \Omega \, cm^2/\rho \tag{8}$$

with $\rho$ in ohm centimeters. Values of $l_e$ are tabulated in Table I.

The induced-current method for resistivity measurements offers exciting possibilities which might otherwise not be feasible. In addition to resistivity measurements on cavities and other unwieldy specimens, superconducting parameters such as $\kappa$, $H_{c1}$, and $H_{c2}$ may also be obtained. After vacuum firing, the material near the (nearly identical) inner and outer cavity surfaces may be relatively more pure than in the bulk, and may therefore show higher values of $H$, than the bulk. From our measured values of $H_{c1}$, and the known correlation of k, $H_c$, $H_{c1}$, and $\kappa$, we may deduce a value for $\kappa$ in the near-surface layer. From either the $H_{c2}$, or resistivity measurements, another value of $\kappa$ may be determined for the bulk. From the two independently determined values of $\kappa$, our apparatus can, in principle, contribute to the knowledge of the radial impurity distribution.

**Acknowledgments**

We wish to acknowledge our appreciation to the SLAC rf superconducting group for making the Q and $H_p'$ measurements, to Earl Hoyt for vacuum firing the Nb, and to Werner Schulz for construction of the pickup coil.

*Work supported by the U. S. Atomic Energy Commission.